\documentclass{jpp}
\usepackage{graphicx}
\usepackage{indentfirst}
\usepackage[utf8]{inputenc}
\usepackage[T1]{fontenc}
\usepackage{amsmath}
\usepackage{xcolor}
\usepackage{dcolumn}% Align table columns on decimal point
\usepackage{bm}% bold math
\usepackage[utf8]{inputenc}
\usepackage[T1]{fontenc}
\usepackage{mathptmx}
\usepackage{etoolbox}

\usepackage{color}

\shorttitle{Laser pulse focusing and energetic electron generation by magnetized  plasma lens}
\shortauthor{Trishul Dhalia, Rohit Juneja and Amita Das}
\title{Laser pulse focusing and energetic electron generation by magnetized  plasma lens}

\author{Trishul Dhalia\aff{1}
  \corresp{\email{trishuldhalia@gmail.com}},
  Rohit Juneja\aff{1},
Amita Das\aff{1}
\corresp{\email{amita@iitd.ac.in}} }

\affiliation{\aff{1} Department of Physics, Indian Institute of Technology Delhi, Hauz Khas, New Delhi-110016, India }
\vspace{10pt}

\begin{document} 

\maketitle

\begin{abstract}
An efficient novel mechanism of laser pulse focusing with the help of a shaped underdense plasma target immersed in an inhomogeneous magnetic field has been demonstrated. These studies have been carried out with the help of 2-D Particle-In-Cell (PIC) simulation  employing the OSIRIS 4.0 platform. It is shown that the divergent magnetic field profile compresses the EM wave pulse in the transverse direction.  A comparative investigation with plane and lens-shaped  plasma geometries  has also been conducted to find an optimal configuration for focusing the laser at the desirable location. Furthermore, it is also demonstrated that when the electron cyclotron resonance (ECR) layer is placed at a suitable location where the laser is focused, a highly energetic electron beam gets generated.

%Harmonic generations with magnetized plasma providesan efficient mechanism to produce high-frequency coherent Electromagnetic (EM) waves in a controlled manner \citep{maity2021harmonic,fu2008harmonic} by using a low frequency incident EM pulse on an overdense plasmamedium. In the present study we have carried out Particle – In – Cell simulations using EPOCH framework to show that harmonic generation also occurs in a configuration for which the external magnetic field is appliedparallel to the EM wave propagation (i.e. RL mode geometry). 
\end{abstract}

\section{Introduction}
The guiding and focusing of intense electromagnetic waves in plasma channels \citep{ehrlich1996guiding,hubbard2002high,esarey1997self} have a wide range of applications in fusion plasma, harmonic generation, and laser-plasma accelerators \citep{kaw2017nonlinear, das2020laser,nishida1987high,ganguli2016development,gibbon1996short,pukhov1999particle}. Many of these applications require tight focusing of high-intensity lasers. In vacuum, a high-intensity laser can propagate up to a few Rayleigh lengths $(X_R=\pi r_0^2/\lambda)$ without diffraction, where $\lambda$ is the laser wavelength and $r_0$ is radius of the spot size of the laser. It is well known that self-focusing in underdense plasma occurs when a relativistically intense laser pulse propagates through it \citep{sprangle1987relativistic}. For a sufficiently high-power laser, $P$ $\geq$ $P_N$, the laser becomes self-focused in the plasma medium. Here $P_N$ is the critical power required for nonlinear self-focusing. Physically, the ponderomotive force of a focused EM wave pushes the electrons out of the high-intensity region.
%which leads to density compression/depression in the plasma. 
The change in density modifies the refractive index of the medium, resulting in the self-focusing of the high-intensity laser. However, this process becomes inefficient if the laser intensity is in the nonrelativistic regime or if the laser power is less than the critical power. Relativistic self-focusing over longer distances can be improved by choosing a preformed density profile. To overcome the difficulty of periodic focusing/defocusing, a localized upward plasma density ramp has been proposed \citep{gupta2007additional,gupta2007plasma}. Recent work on long-chirped pulse compression using a preformed density ramp has also suggested the possibility of reaching exawatt or zetawatt levels \citep{hur2023laser}. At these intensities, the laser pulse will be self-focused. However, in a preformed plasma density profile, there are energy losses resulting from Raman backscatter (RBS) at the quarter-critical density $(n_c/4)$ \citep{hur2005electron,hur2023laser}. These energy losses could be minimized by choosing a sharp density cut-off at $(n_c/4)$. Thus, the conventional self-focusing of a laser pulse using a plasma medium has limitations, such as the requirement of very high intensity and energy losses due to parametric processes.

%The shape of the plasma channel can also play an important role in self-focusing. Ren \textit{et al}. have demonstrated the compression and focusing of short laser pulses using a thin plasma lens \citep{ren2001compressing,hubbard2002high}.\\

The interaction of intense laser pulses with magnetized plasma has recently attracted considerable interest. The laser frequencies being high, the requirement for a magnetic field to elicit a magnetized response from plasma at the laser frequency is quite high. However,  for the low-frequency $CO_2$ lasers,  a magnetic field of the order of $\sim kT$ suffices.  It is now technologically possible to generate magnetic fields of this strength in laboratory environments \citep{nakamura2018record}. Similarly, with the availability of high-power microwave pulses (HPMs $P> 1GW$) \citep{nevins1987nonlinear,allen1994nonlinear,xiao2020efficient} interaction of  electromagnetic (EM)  waves with magnetized plasma can be explored in the nonlinear regime. Microwave frequencies are in the $\sim$ GHz range, which requires $\sim 0.1-1 T$ order magnetic field to elicit a magnetized response from electrons.  In simulations, both these regimes can be explored in terms of  normalized parameters of the plasma and the EM source. There are three possible distinct geometries  termed as the RL, X, and O-mode configurations.  In the RL mode the EM wave propagation is parallel to the applied magnetic field; in X and O modes it is perpendicular. While in X mode the laser magnetic field  is parallel to the applied field; in O mode it is perpendicular to the same.  \citep{swanson2020plasma,stix1992waves,chen1984introduction}. Lately,  a number of  simulation studies have been carried out for  these  configurations addressing a wide variety of issues.  For instance, study of harmonic generation \citep{dhalia2023harmonic,maity2021harmonic}, localized energy absorption \citep{PhysRevE.105.055209,vashistha2022localized,juneja2024enhanced,PhysRevE.110.065213}, ion heating \citep{Juneja_2023,vashistha2020new}, parametric excitation \citep{goswami2022observations} and magnetic transparency of plasma medium \citep{mandal2021electromagnetic} and controlled electron beam generation \citep{choudhary2023controlling} have been studied.   

In this work, we study the role of magnetic field and shaped plasma targets in self-focusing of electromagnetic waves using Particle - In - Cell simulations. Analytical investigation of the role of a homogeneous magnetic field on the self-focusing of an intense laser beam in the presence of a transverse as well as axial magnetic field has been carried out by  \citep{jha2006self, jha2007spot}.
%In the context of hot magnetized plasma, the self-focusing of circularly polarized waves has been studied  analytically by \cite{abedi2017relativistic}.
Under a constant applied external magnetic field along the propagation axis, self-focusing of the circularly polarized wave is reported in \citep{molavi2024investigation}. A constant magnetic field can induce self-focusing at the discontinuity of the refractive index, which will occur at the vacuum plasma interface of a homogeneous plasma. However, in real scenario the magnetic field may not necessarily be constant throughout the plasma. We explore the impact of the spatial variation of the magnetic field on self-focusing. For this purpose, a two-dimensional Particle-in-cell simulation was carried out.   Through our simulations, we illustrate a novel mechanism that enables the self-focusing of a high-intensity electromagnetic pulse in the presence of an axially diverging magnetic field. We have chosen a homogeneous, fully ionized underdense plasma for our study. In addition, we have also studied the impact of the plasma target shaping. Both a slab and a plasma target in the form of a convex lens has been considered. 
%Although the shape might not be as ideal in experiments as we have designed but these types of geometry can be replicated in micro-droplet experiments \cite{kurilovich2016plasma} as well as curved plasma channels of low density plasmas\cite{ehrlich1996guiding,luo2018multistage,zhu2023experimental,hubbard2002high}. Ren \textit{et al}. have demonstrated the compression and focusing of short laser pulses using a thin plasma lens \cite{ren2001compressing}. The chosen magnetic field geometry is not very complex and can be easily generated with the help of a single permanent magnet or an electromagnet. The magnetic field values for laser-plasma and microwave plasma interactions are mentioned in the next section. The idea behind the approach is that electrons are trapped along the plane perpendicular to the applied external magnetic field's direction. However, under the influence of high-intensity EM wave pulse, electrons will quiver in that perpendicular plane, thus creating a density depression along the axial direction and compression along the transverse direction. This density fluctuation will modify the plasma refractive index, and the incoming EM wave will become self-focused along the axial direction. \\

 The paper is organized as follows. Section \ref{sec:simulation} describes the simulation geometry and the choice of parameters. In section \ref{sec: focusing} we carry out a comparative study of self-focusing in three different configurations: a) magnetized lens, b) unmagnetized lens c) magnetized slab. Subsection \ref{sec:theory} presents analytical calculations for spot size evolution in the chosen geometry and describes the effect of polarization on self-focusing. In section \ref{sec:electron_beam} the energetic electron beam generation has been demonstrated. In section \ref{sec:summary} we summarize our findings and conclude. 

%%%%%%%%%%%%%%%%%%%%%%%%%%%%%%%%%%%%%%%%%%%%%%%%%%%%%%%%%%%%%%%%%%%%%%%%%%%%%%%%%%%%%%%%%%%%%%%%%%%%%%%%%%%%%%%%%%%%%%%%%%
\begin{figure*}
  \centering
  \includegraphics[width=1\linewidth]{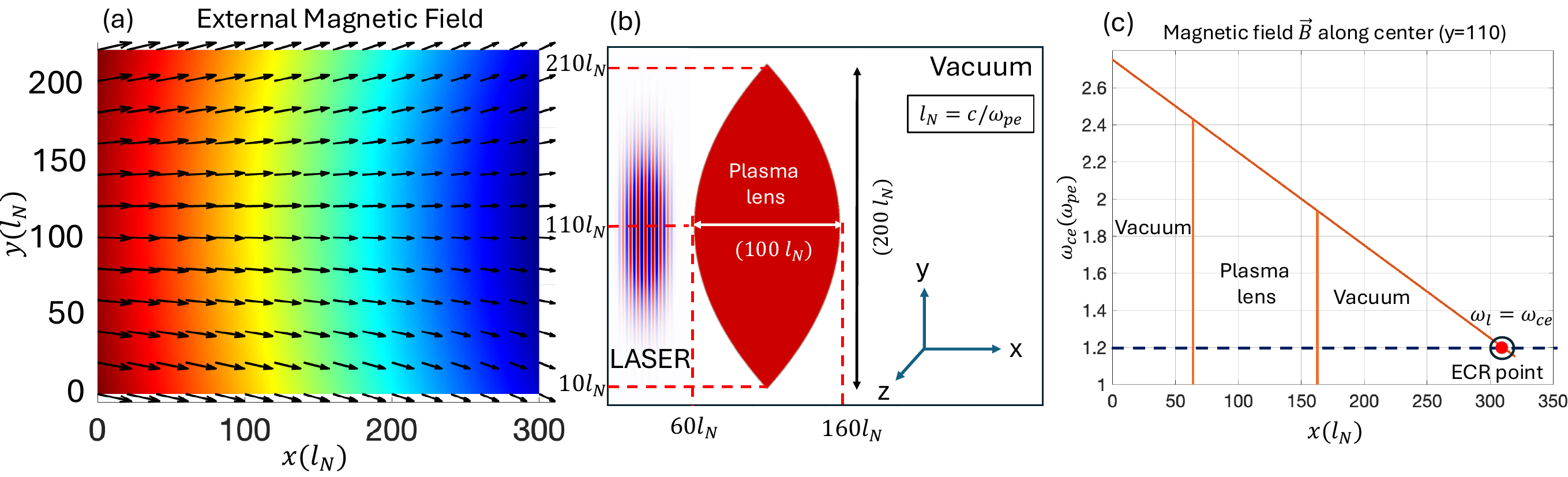}% Images in 100% size
  \caption{Figure here demonstrates the schematic representation (not to scale) of the geometry chosen for simulation. Figure (a) shows the magnetic field profile applied in simulation box. Figure (b) shows the shape of plasma lens and pulse profile of incident EM wave. Figure (c) shows plot of external magnetic field  in terms of $\omega_{ce} =eB_0/m_e$ along center axis (y=110$l_N$) with respect to x.}
\label{fig:schematic}
\end{figure*}

\begin{table*}
  \begin{center}
\def~{\hphantom{0}}
  \begin{tabular} {cccc}
  \hline
      $$\textbf{ Parameters}$$       & $$ \textbf{ Normalized values}$$   &   $$  \textbf{ Microwave System}$$ & $$  \textbf{Laser System}$$  \\[4pt]
      
    Frequency($\omega_{EM}$)  & 1.2$\omega_{pe} $ & $ 4.2\times 10^{10}rad \ s^{-1}$ & $0.2\times 10^{15}rad \ s^{-1}$ \\ 
  
    Wavelength($\lambda_{EM}$) & 5.23$c/\omega_{pe}$ &  44.88 $mm$  & 9.42  $\mu m $\\
  
    Intensity ($I_0$)           & $a_0=0.08$        &  $4.34\times 10^{6} W \ cm^{-2}$ & $9.86\times 10^{13} W \ cm^{-2}$\\
    \hline
                         \multicolumn{4}{c}{\textbf{Plasma Parameters}} \\
    
      Density$(n_{e,i})$   & 1  & $3.85\times 10^{11} cm^{-3}$ &  $4.47\times 10^{19}  cm^{-3}$\\
      
    ($\omega_{pe})$   & 1 & $3.5\times10^{10}rad \ s^{-1}$ & $3.77\times 10^{14} rad \ s^{-1}$ \\
     
      ($c/\omega_{pe})$ & 1 & 8.57 mm & $0.79 \mu m$\\
      \hline
                    \multicolumn{4}{c}{\textbf{External Magnetic Field }} \\
  
    $B_0$       &   2.2 $m_ec\omega_{pe}e^{-1}$ &        $\sim$ 0.44 T   & $ \sim$  4.71 kT  \\
    \hline
  \end{tabular}
  \caption{Simulation parameters are shown here in normalized as well as in corresponding SI units}
  \label{tab:parameters}
  \end{center}
\end{table*}

\section{Simulation Details}\label{sec:simulation}
\indent The simulation geometry is depicted in Fig. (\ref{fig:schematic}). For our study 2-D particle-in-cell (PIC) code has been employed using the OSIRIS 4.0 platform \citep{hemker2000particle, fonseca2002osiris, fonseca2008one}. The choice of parameters for this study has been provided in Table-\ref{tab:parameters}. The macroparticles chosen in our simulations are $4\times 4$=16 particles per cell. The spatial and temporal resolution has been taken as $dx,dy=0.1c/\omega_{pe}$ and $dt=0.02\omega_{pe}^{-1}$, respectively. The length and time scales are normalized by $l_N \rightarrow c/\omega_{pe} $ and $t_N\rightarrow\omega_{pe}^{-1}$. The electric and magnetic fields are normalized by $E_N=B_N=m_ec\omega_{pe}e^{-1}$ where $m_e$ and $e$ are electron mass and charge, respectively. The extent of the simulation box has been chosen 300$l_N$$\times$200$l_N$ for this study. We have considered a fully ionized electron-ion plasma with homogeneous density. The mass ratio of ion to electron is taken $1836$. We have performed simulations for three cases in our study for comparison, a) magnetized plasma lens, b) unmagnetized plasma lens, and c) magnetized plasma slab. The homogeneous plasma considered in our study is defined in the cartesian geometry as,
\begin{equation}
    c(y-110)^2 +60 \leq x\leq -c(y-110)^2 +160
\end{equation}
 Here, parameter $c$ determines shape of the plasma. For the plasma lens geometry $c=0.005$ and $c=0$ will give a plasma slab geometry. The plasma region extends from $x=60 l_N$ to $x=160 l_N$ and $y=10 l_N$ to $y=210 l_N$ and the central axis is along $y=110l_N$. The incident EM wave has frequency $\omega_{EM}=1.2\omega_{pe}$. Here $\omega_{pe}$ represents normalized plasma frequency. The spatial profile of the pulse is Gaussian. The Incident laser pulse focused initially at the vacuum plasma interface at $x=60l_N$. The pulse duration of the incoming wave is $50 \omega_{pe}^{-1}$ and the transverse initial spot-size has a diameter of $80 l_N$. The vacuum diffraction length of the incoming pulse is $k_0r_0^2/2=960l_N$. Thus within entire simulation box length $(300l_N)$ incoming pulse's spot-size remains unchanged until it interacts with plasma. The spot-size initially is taken considerably smaller than the transverse extent $(220l_N)$ of the plasma to minimize the effect of spherical aberration \citep{li2022influence}. 
% The external magnetic field applied in the simulation has been intentionally chosen to be of a divergent nature. 
 The profile of the external magnetic field has the following form, 
 \begin{equation}
     \vec{B}_{ext}= [2.2 - \delta(x-110)]\hat{i}+ [\delta(y-110)]\hat{j} B_N
 \end{equation}

 The chosen external magnetic field profile has a divergent nature from the center line $y=110 l_N$, which satisfies the required condition of $\vec{\nabla}.\vec{B}=0$. Here, the value of parameter $(\delta =0.005)$ is chosen such that the associated gyrofrequency $(\omega_{ce})$ due to the external magnetic field at the end of the target plasma stays larger than the incident EM wave frequency $[\omega_{ce}(x=160 l_N) >\omega_{EM}]$ as shown in figure \ref{fig:schematic}(c). This ensures that the group velocity of the EM wave, within the plasma stays finite. The simulation run time was $300\omega_{pe}^{-1}$. During this time, ion response is negligible compared to electrons and they merely act as a neutralizing background \citep{PhysRevE.110.065213,PhysRevE.105.055209}. Thus, the driving source of focusing for incoming EM wave pulse is associated with the dynamics of electrons, which being lighter, respond at such a fast time scale. However, it should be noted that both the electron and ion dynamics have been considered in simulation.
 The normalized vector potential of EM wave is taken $a_0=0.08$ to stay within a non-relativistic regime. This intensity has been chosen such that the self-focusing due to relativistic nonlinearity can be avoided and we can have a direct comparison of unmagnetized  and magnetized plasma. The chosen normalized parameters can  represent both  a laser-plasma interaction as well as  also  microwave-plasma interaction by using appropriate normalization factors. This has been illustrated in Table -\ref{tab:parameters}. \\
%  The external magnetic field required for microwave-plasma study is given as, 
% \begin{equation}
%      \vec{B}_{ext}= [0.2 -0.00045(x-110)]\hat{i}+ [0.00045(y-110)]\hat{j} \ \text{T}
%  \end{equation}
%  For laser-plasma interaction, the external magnetic field required is given as, 
% \begin{equation}
%      \vec{B}_{ext}= [24.49 -0.055(x-110)]\hat{i}+ [0.055(y-110)]\hat{j} \ \text{kT}
%  \end{equation}
% Thus, we only need $\sim 0.2 T$ order of magnetic field to conduct a microwave experiment, whereas, $\sim 25 kT$ order of magnetic field is required for an 800 nm laser to undergo focusing.  \textcolor{red}{Amita: how ? also will it not be a good idea to talk about the microwave system at the end in one paragraphedited up to here }
 \indent The microwave system features a pulse, high power microwave of spot size 68.56 cm and  frequency 6.7 GHz and power of approximately 16 GW incident on a plasma lens of height 171.4 cm and width of 85.7 cm \citep{nevins1987nonlinear,allen1994nonlinear}. The external magnetic field required  here is of order $\sim 0.44 T$ or 4.4 kilo gauss, which can be generated through a simple permanent magnet or an electromagnet. Similarly, a laser system will feature a  moderate power 0.03 TW, $\text{CO}_2$ laser of spotsize 63.2 $\mu m$ , incident on a plasma lens of height $158 \mu m$ and width of $79 \mu m$. The external magnetic field required is of order $4.71 kT$ or 47.1 mega gauss \citep{choudhary2025generation,hao2025generation}.   
%%%%%%%%%%%%%%%%%%%%%%%%%%%%%%%%%%%%%%%%%%%%%%%%%%%%%%%%%%%%%%%%%%%%%%%%%%%%%%%%%%%%%%%%%%%%%%%%%%%%%%%%%%%%%%%%%%%%%%%%%%
%%%%%%%%%%%%%%%%%%%%%%%%%%%%%%%%%%%%%%%%%%%%%%%%%%%%%%%%%%%%%%%%%%%%%%%%%%%%%%%%%%%%%%%%%%%%%%%%%%%%%%%%%%%%%%%%%%%%%%%%%%%%
\begin{figure*}
    \centering
    \includegraphics[width=1\linewidth]{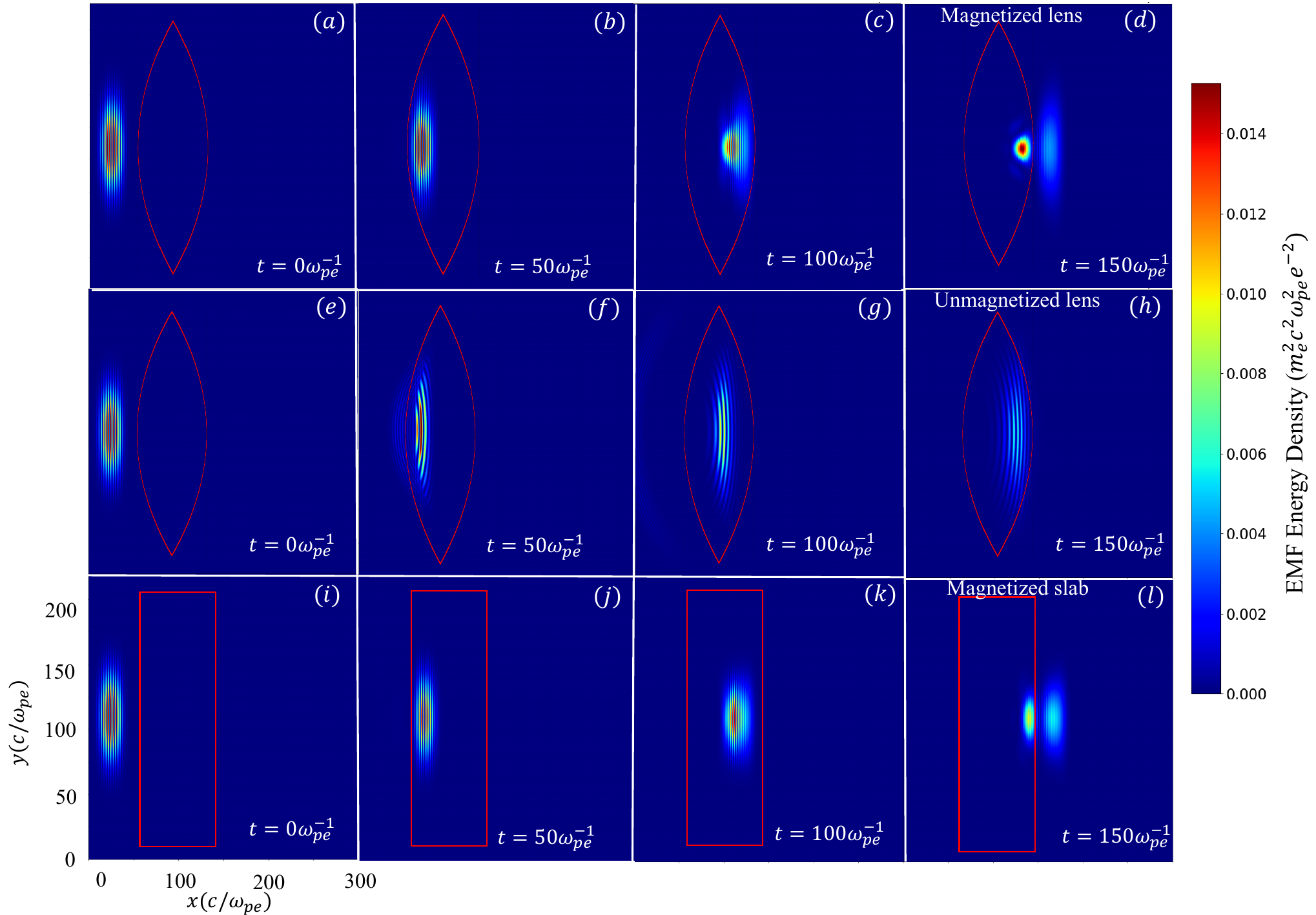}
    \caption{Time evolution of EMF energy density in the presence of i) Magnetized plasma lens (first row), 2) Unmagnetized plasma lens (second row) and 3) Magnetized plasma slab (third row) has been presented.}
    \label{fig:time_evolution_Emf}
\end{figure*}
\begin{figure*}
    \centering    \includegraphics[width=1\linewidth]{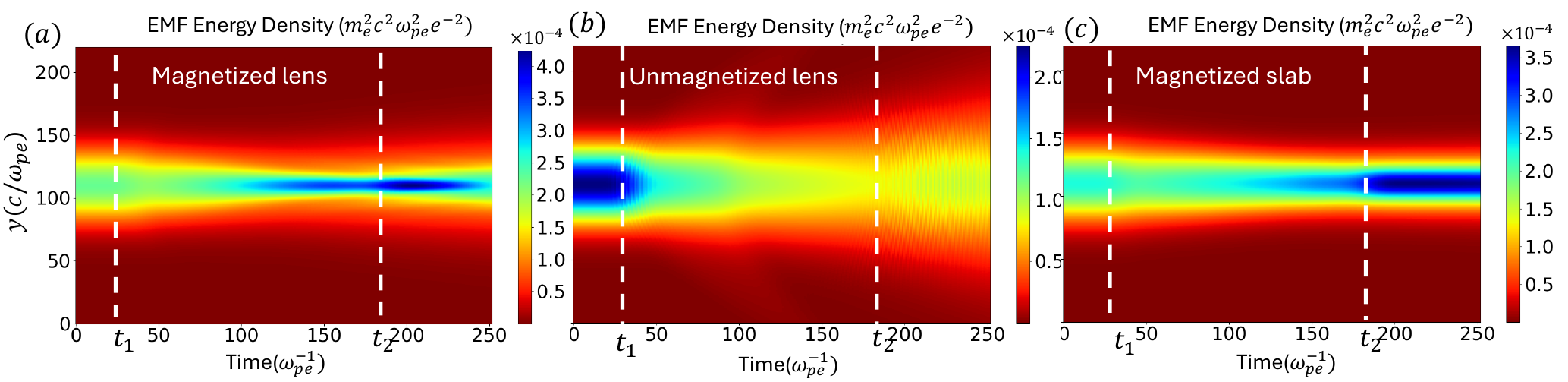}
    \caption{Figure demonstrates x-averaged EMF energy density with time and y-direction for a) Unmagnetized lens b) Magnetized lens c) Magnetized slab. Here, $t_1$ and $t_2$ denotes time at which wave pulse enter and leave the plasma boundary respectively.}
    \label{fig:focussing_pulse}
\end{figure*}
\begin{figure}
    \centering   
    \includegraphics[width=0.5\linewidth]{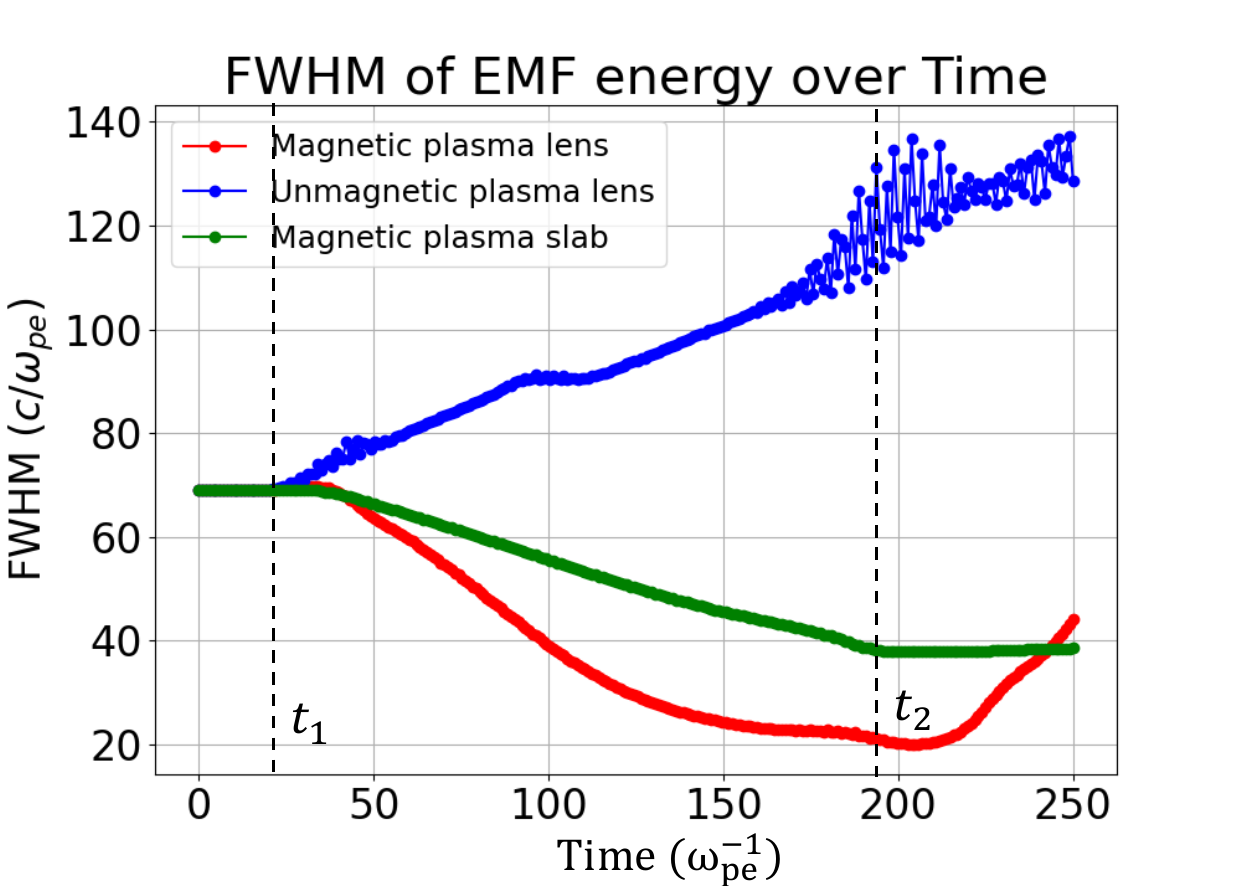}
    \caption{Figure shows the time evolution of full width half maxima (FWHM) of incident wave with a) Unmagnetized lens b) Magnetized lens c) Magnetized slab }
    \label{fig:spot_size}
\end{figure}
\begin{figure*}
  \centering
  \includegraphics[width=1\linewidth]{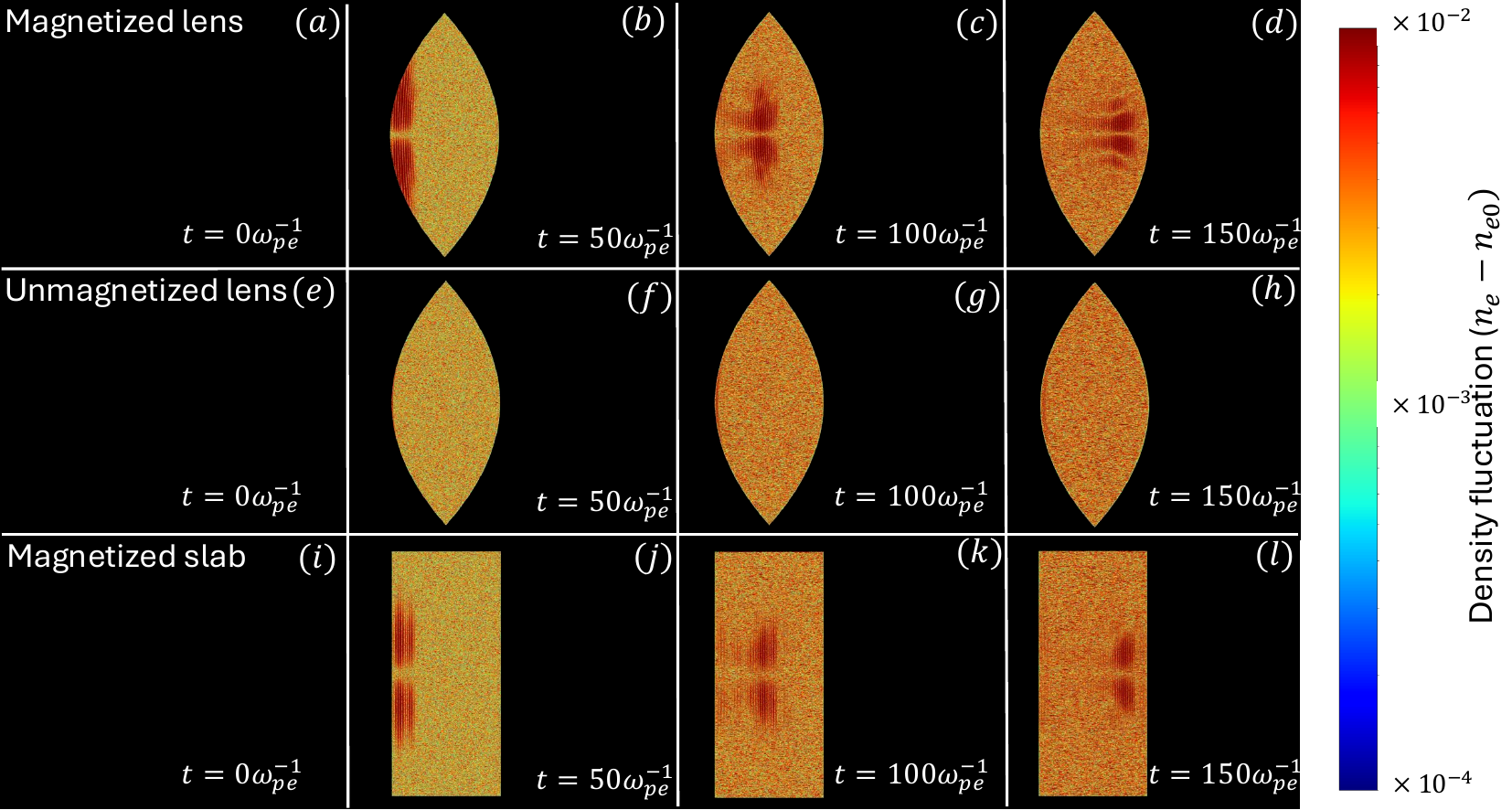}% Images in 100% size
  \caption{Time evolution of net density fluctuation in electron density for three cases of magnetized lens, unmagnetized lens and magnetized slab.}
\label{fig:density_fluctuation}
\end{figure*}
% \begin{figure*}[!ht]
%   \centering
%   \includegraphics[width=14cm]{particle_trajectory_all.pdf}% Images in 100% size
%   \caption{Particle trajectory of 1000 randomly chosen particle from the plasma regions. Insets shows individual particle trajectory at shown location with time. The red arrows show direction of external magnetic field.}
% \label{fig:particle_trajectory_all}
% \end{figure*}
\begin{figure}
  \centering
  \includegraphics[width=8cm]{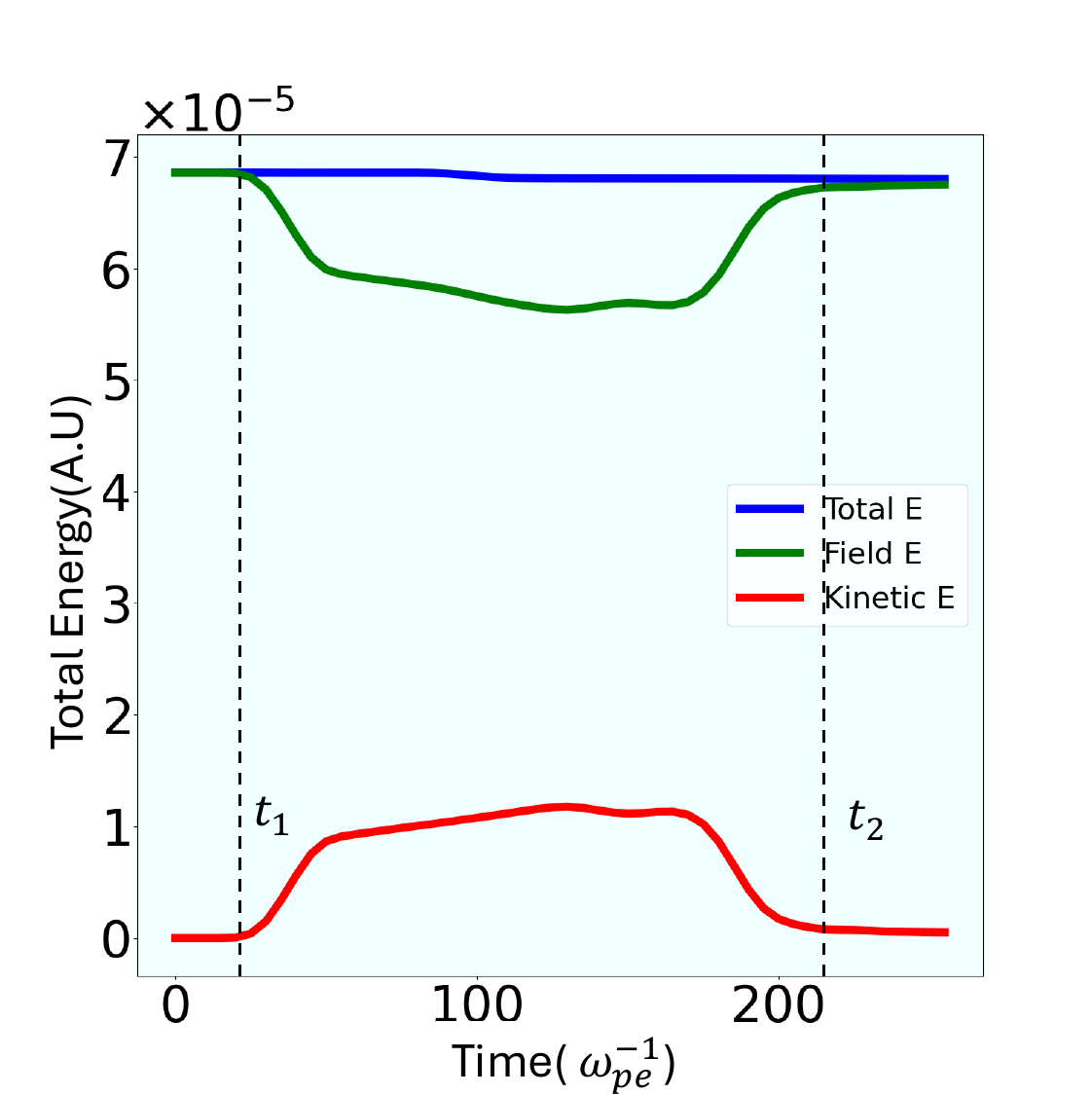}% Images in 100% size
  \caption{Time evolution of Total energy, EMF energy and kinetic energy of electrons for the case of magnetic lens.}
\label{fig:Energy_lens}
\end{figure}
\section{Magnetic field induced focusing }\label{sec: focusing}
An incoming linearly polarized EM wave pulse   typically breaks up into right (RCP)  and left (LCP)  circularly polarized waves while propagating parallel to an applied external magnetic field in a plasma medium. The separation occurs as R and L-modes have different group velocities. This propagation is depicted in figure \ref{fig:time_evolution_Emf} by plotting the electromagnetic field energy density at various times as the pulse propagates through three different configurations of  finite sized  plasma target.   The first corresponds to a  target in the form of a  plasma lens immersed in magnetic field, the second has the same target profile and no   magnetic field is applied. The third corresponds to a plasma slab immersed in a magnetic field.

\indent At the vacuum plasma interface, the incoming wave pulse breaks into RCP and LCP wave pulse for the case of magnetized lens. Figure \ref{fig:time_evolution_Emf}(b,c,d) shows that spot size of slow RCP wave is decreasing.  On the other hand the LCP wave pulse  does not get focused instead it shows divergence. In comparison, in the next row snapshots of the EMF energy density for an unmagnetized lens have also been shown at various times. The unmagnetized lens does not support slow and fast modes and merely follows the dispersion relation of unmagnetized underdense plasma. The  EM wave passes through it and does not show focusing  inside  the plasma. 
The third case is of a magnetized plasma slab for which snapshots have been presented in the third row of figure \ref{fig:time_evolution_Emf}. In this scenario the wave again divides into slow RCP and fast LCP waves. However the observation indicates that the focusing of the RCP wave in this case compared to the first case is considerably weak.

\indent Figure \ref{fig:focussing_pulse} shows the space -time ($y$ vs. $t$)  plots of EMF energy averaged over x-direction for all the three cases.  Here, $t_1$ and $t_2$ indicates the times when EM wave start interacting with plasma and exit the plasma surface. It is evident from these plots that for unmagnetized case, pulse EMF energy density was maximum around $\sim(2\times 10^{-4} m_ec\omega_{pe}e^{-1})$ initially, (before interaction) and it diverges in the transverse direction  as it   interacts with plasma. Thus, no focusing has been observed in this scenario. On the other hand, for the case of the magnetized lens, laser spot size acquires a minimum spot and maximum EMF energy density $\sim(4\times 10^{-4} m_ec\omega_{pe}e^{-1})$ around double of it's initial  energy after the pulse has interacted with plasma at time near $200 \omega_{pe}^{-1}$. This is also interesting to notice that afterwards pulse again defocuses and goes out of the simulation box. For the case of a plasma slab, again wave pulse acquires a minimum spot and the peak EMF energy density $\sim(3.5\times 10^{-4} m_ec\omega_{pe}e^{-1})$. However, this focusing is not as tight as that for the plasma lens geometry of the target. Here, the spot size does not defocus after pulse leaves the plasma boundary and remains constant with time.

\indent We have also provided a comparison of the  evolution of  FWHM (full width at half maxima) for each of the three cases.  This has been plotted  in figure \ref{fig:spot_size}. While for unmagnetized lens spot size shows divergence from the beginning itself, in the case of magnetized lens, it gets focused by a factor of nearly $1/3.5$ and attains the value of $(\sim 20 c/\omega_{pe})$. Thereafter,  coming outside the plasma surface it again starts defocusing. For the  slab geometry the focusing is weaker and it attains a value of   $38 c/\omega_{pe}$, which is a reduction by a factor of $1/2$.  It, therefore, appears that the  convexity of the plasma profile plays a role for tighter focusing. 
%tries  to focus the wave pulse at the focal point and provides maximum compression of spot size. While in the case of plasma slab compression is only due to magnetic field inhomogeneity. In this case  only a linear decrease in FWHM is observed from $t_1$ to $t_2$.

Figure \ref{fig:density_fluctuation} shows the comparison of plots for the electron density fluctuation for these three cases. It should be noted that a spatially inhomogeneous electron density distribution results for the two cases in which the magnetic field is present. The electron density bunches  changes the refractive index and is responsible for efficiently focusing the spotsize of the EM wavepulse effectively in these cases.

These studies  thus  illustrate  that both the shape of plasma  and the profile of the magnetic field play crucial roles in EM wave pulse focusing. It should be noted that this is a  novel approach. The  focusing here does not require relativistic intensities of EM wave-pulse. In fact even a  moderately intense pulse can be focused at a desired location.  Another important feature is that almost no energy is dissipated into plasma during this interaction. Figure \ref{fig:Energy_lens} shows time evolution of total energy, field energy and kinetic energy for the case of magnetic lens. From this plot we observe that the total energy remains almost constant. There is a conversion from field to kinetic energy in the beginning of the interaction. Thereafter, however, the kinetic energy again converts back into field energy. 
%This can be clearly seen that as the wave-pulse interacts with plasma, the kinetic energy of particles rises, because of quiver motion of electrons under the influence of EM wave. As soon as the EM wave leaves the plasma lens, kinetic energy of electron becomes negligible and field energy nearly becomes the net total energy and leaves the simulation box at later times. During the interaction time, heavier species ions are not able to respond, and the lighter species electrons come to rest after the interaction. This means no parametric processes can occur and the mechanism does not excite any kind of parasitic instability.   
%%%%%%%%%%%%%%%%%%%%%%%%%%%%%%%%%%%%%%%%%%%%%%%%%%%%%%%%%%%%%%%%%%%%%%%%%%%%%%%%%%%%%%%%%%%%%%%%%%%%%%%%%%%%%%%%%%%%%%%%%%
We now try to understand the contrasting response of focusing  displayed by the  LCP and the RCP EM waves in the next subsection.

\begin{figure}
     \centering    
     \includegraphics[width=1\linewidth]{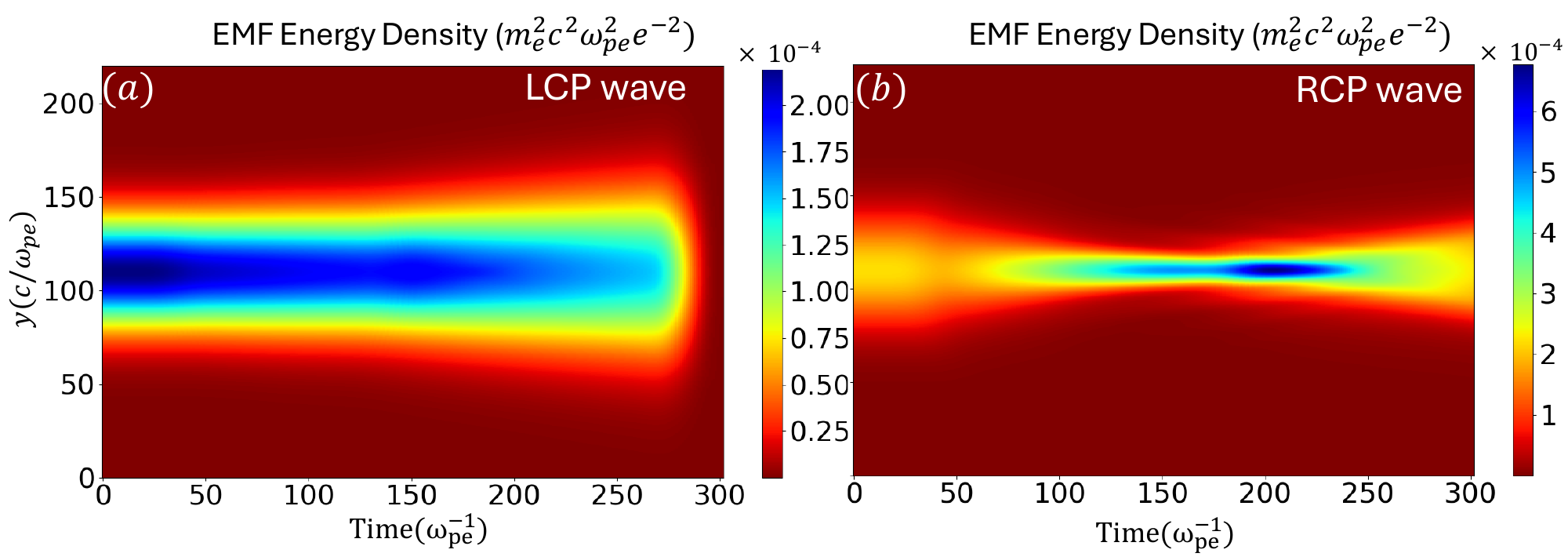}
    \caption{Figures shows x-averaged time (t)-space (y) evolution of a) LCP wave pulse b) RCP wave pulse in the presence of magnetized plasma lens}
     \label{fig:polarization_effect}
 \end{figure}
 \begin{figure*}
    \centering   
    \includegraphics[width=1\linewidth]{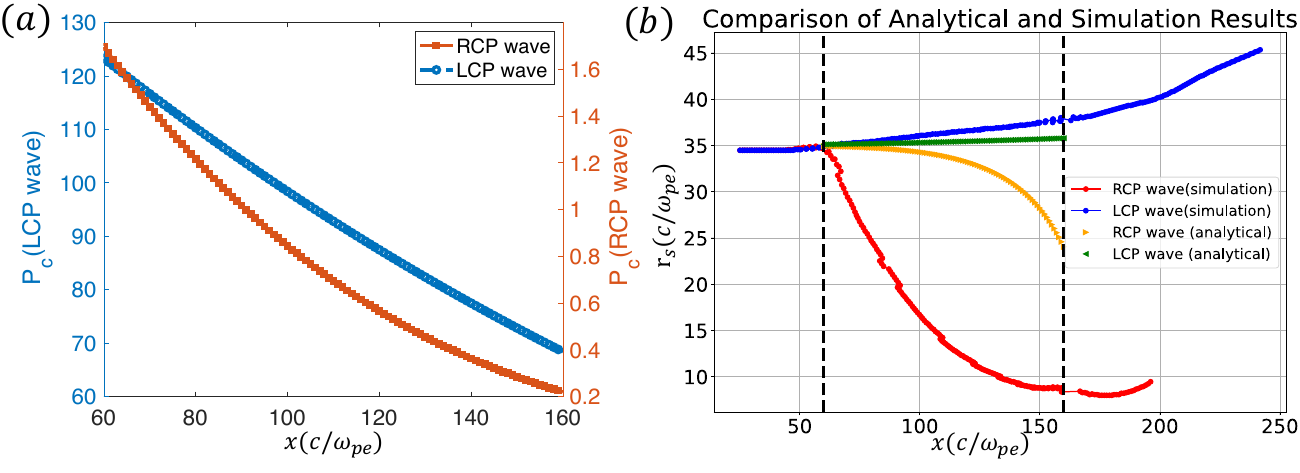}
    \caption{Figures (a) shows critical power of RCP and LCP wave in the magnetized plasma along x-direction and figure (b)  demonstrates evolution of beam waist $r_s$ of RCP and LCP wave in the magnetized plasma along propagation direction from analytical expression and simulation data. }
    \label{fig:power+focusing}
\end{figure*}

\subsection{Interpretation  }\label{sec:theory}
We have also carried out simulations for an incident wave with a definite RCP and LCP form of polarization. Here the waves do not break up in two pulses as expected. 
However, only the RCP wave shows focusing whereas the wave pulse with LCP shows divergence as seen from figure \ref{fig:polarization_effect}. This distinctive behavior of focal spot evolution for the two polarizations have been understood  analytically by applying a local approximation to the theory put forth for the homogeneous magnetic field by \citep{jha2006self,jha2007spot}. In there study the source dependent expansion (SDE) method \citep{esarey1997self}, is used to study the evolution of the laser spot. The expression for laser spot evolution in the presence of an axial magnetic field is given as , 
                    \begin{equation}\centering
                                 \frac{r_s^2}{r_0^2}= 1+ (1 -\frac{P}{P_{c}})\frac{x^2}{X_R^2}
                    \end{equation}
Here $r_s$ is the instantaneous beam waist and $r_0$ is the initial beam waist. P is the power of EM wave pulse given as, $P=(a_0^2r_0^2e^2/4k_0^2c^5m^2)$ and 
\begin{equation}\label{crit_power}
    P_{c} = \frac{k_0^2c^5m^2 }{2k_{p0}^2e^2S}
\end{equation} 
is the critical power required for self focusing. Here $k_0$ and $k_{p0}= 4\pi e^2 n_0/mc^2$ are wave numbers of the EM wave and wave numbers association with the characteristic plasma frequency respectively. The nonlinear parameter induced by the external magnetic field in plasma for the RCP $(\sigma=1)$ and LCP $(\sigma=-1)$  wave is, 
                    \begin{equation}\label{nonlinear}\centering
                            S=\frac{\omega_{EM}^4( \omega_{EM}+\sigma\omega_{ce})^4}{(\omega_{EM}^2-\omega_{ce}^2)^4}
                    \end{equation}
Here, $\omega_{ce}=eB_0/m$. $X_R=(k_0^2r_0^2/2)$  is the Rayleigh length. We have plotted the critical power $P_c$ with the help of equation (\ref{crit_power}) and (\ref{nonlinear})  along the axis for the changing magnetic field in our system. The variation of the critical power with distance $x$ has been shown in figure \ref{fig:power+focusing}$(a)$. It should be noted from the figure that the value of critical power is very high for the LCP wave, whereas for the RCP wave it is quite low. The power of our chosen  EM wave pulse exceeds the critical power for the RCP wave  and hence focusing is observed. 

\indent The role of inhomogeneous magnetic field and the plasma profile provides a much better focusing than what is predicted by the analytical theory of homogeneous magnetic field. This has been illustrated by figure \ref{fig:power+focusing}(b) for which the beam waist evolution in space has been shown for both LCP and RCP waves through simulation and analytical estimate. 

\begin{figure}
    \centering
    \includegraphics[width=1\linewidth]{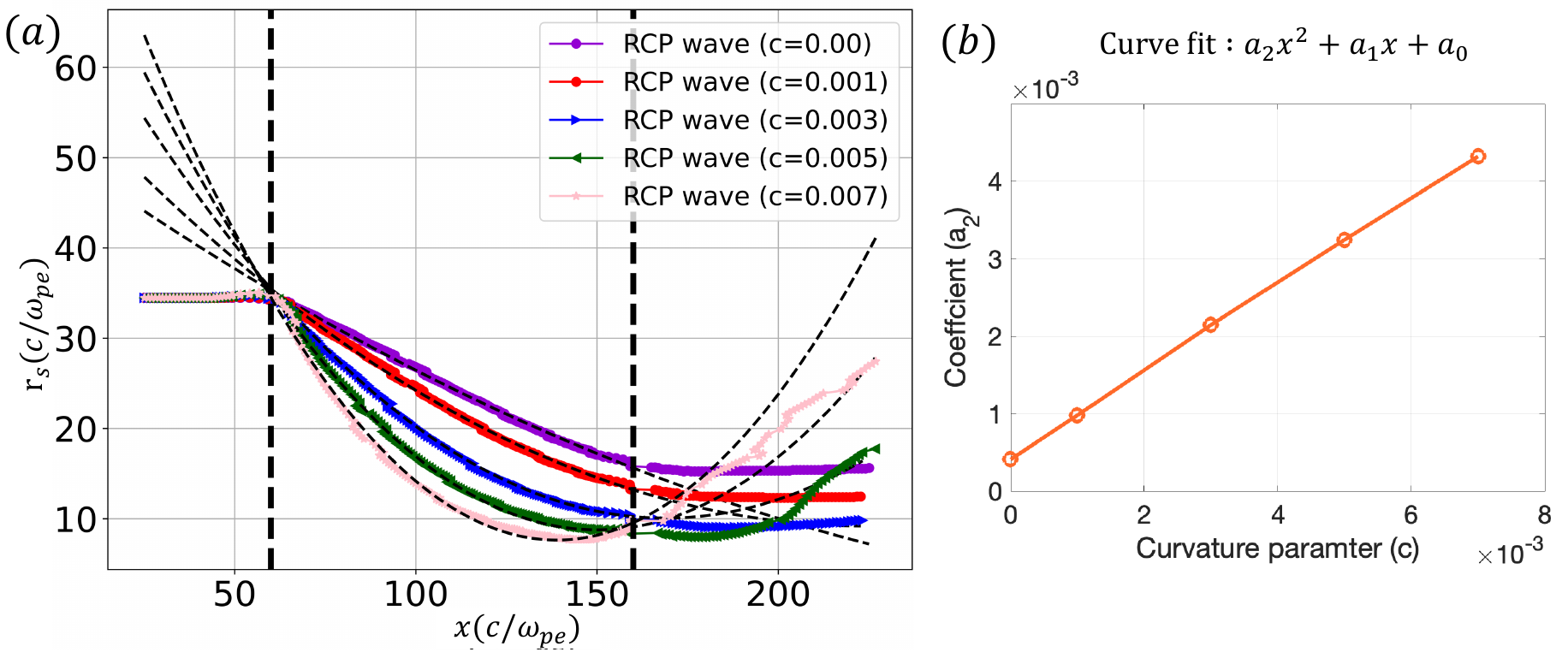}
    \caption{Figure (a) shows the FWHM of incident EM wave for four different curvature of convex plasma geometry and they are fitted with a second order polynomial $a_2x^2+a_1x+a_0$. In figure (b) we plot the coefficient $a_2$ with respect to curvature parameter $c$.   }
    \label{fig:lens effect}
\end{figure}

\subsection{Effect of lens curvature on focusing}
We established in the preceding section that incident RCP wave pulses are strongly focused under the magnetized plasma lens. Here, we want to demonstrate the effect of plasma lens curvature on magnetic field-induced focusing. For this, we have incident RCP waves on five distinct shapes corresponding to the curvature parameter $(c=0,0.001,0.003,0.005,0.007)$ of plasma and kept the magnetic field profile $(\delta=0.005)$ constant in all five cases. The figure \ref{fig:lens effect}(a) shows the evolution of the beam waist $r_s$ of an RCP wave pulse through several plasma lenses. This effectively demonstrates that increasing the curvature of the plasma lens improves focusing. The curves fit well on a second-degree polynomial ($a_2x^2+a_1x+a_0$) within the plasma boundary. In the case of $(c=0.007)$, a minimal spot is observed within the plasma lens, and spot size decreases in a \textsl{quadratic} pattern. It is also worth noting that for a slab geometry $(c=0)$, focusing is lowest and spot size decreases in a \textsl{linear} fashion. Figure \ref{fig:lens effect}(b) demonstrates that as the curvature $c$ grows, so does the coefficient $a_2$ in a linear fashion. Obviously, the curvature parameter should not exceed a particular limit so that the lens height is comparable to or less than the laser spot size; otherwise, spherical aberration will impede the focusing. This concludes that the curvature of the plasma lens provides additional focusing on top of magnetic field-induced focusing.

\begin{figure}
  \centering
  \includegraphics[width=6cm]{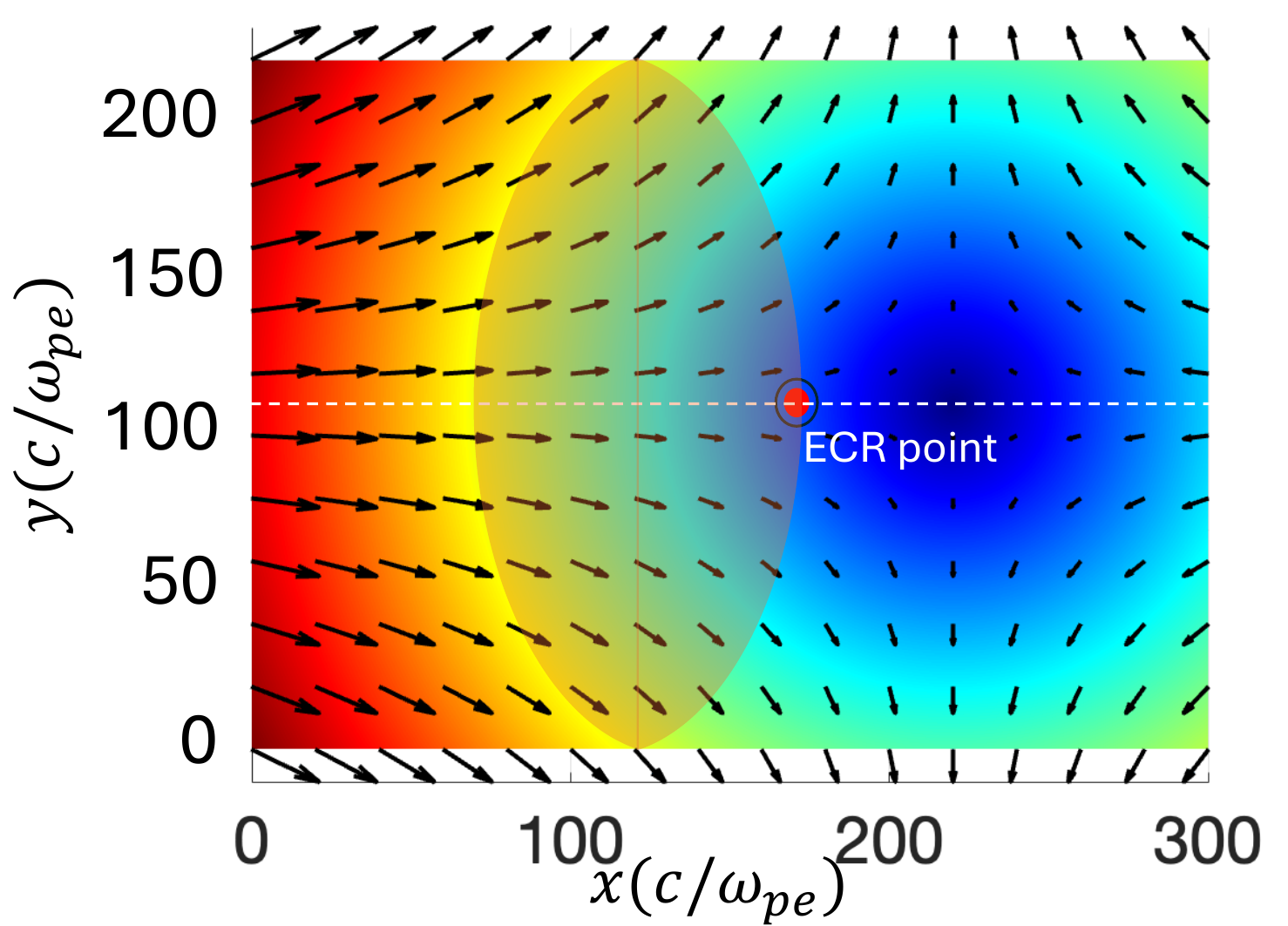}% Images in 100% size
  \caption{Figure demonstrates the plasma lens $(c=0.005)$ situated in a high gradient external magnetic field. The ECR point represent the location at plasma edge where EM wave frequency matches with electron cyclotron frequency.}
\label{fig:schematic_2}
\end{figure}
\begin{figure}
    \centering
    \includegraphics[width=14cm]{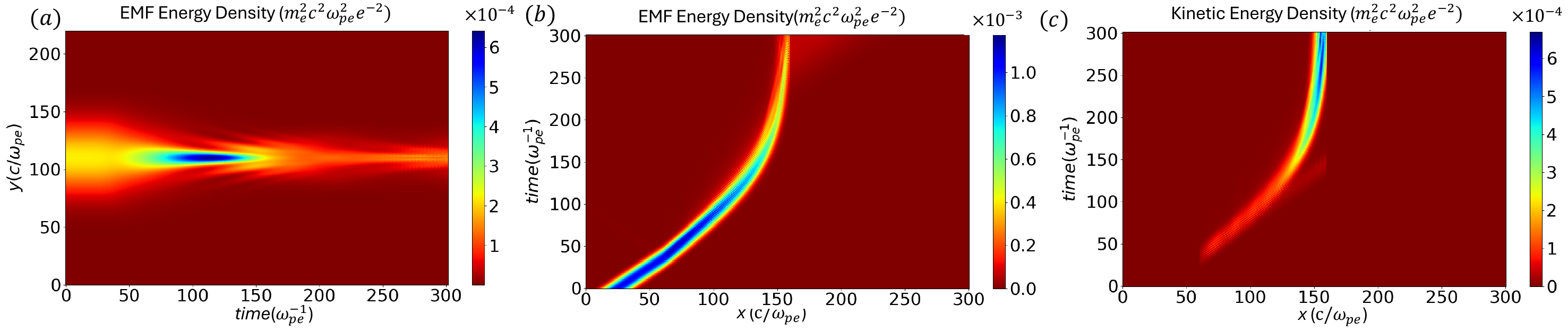}
    \caption{Figure shows space-time plots for EMF energy. $(a)$ y-t plot averaged over x-dimension. $(b)$ t-x plot averaged over y-dimension. $(c)$ t-x plot of Kinetic energy averaged over y-dimension.}
    \label{fig:time_evolution+energy_absorption}
\end{figure} 

\section{Magnetic field induced focused electron beam generation}\label{sec:electron_beam}
In the  previous section, it has been shown that the   RCP wave gets  efficiently focused with the help of a   magnetized plasma lens. We show here that by appropriately tailoring the magnetic field profile with the help of the parameter $\delta$ we can produce energetic electron beam. This happens when the magnetic field profile is tailored to have the ECR resonance lie within the plasma. 
%By increasing the variation parameter $(\delta=0.02)$ of external magnetic field profile, EM wave frequency $(\omega_{EM})$ can be matched with electron cyclotron frequency $(\omega_{ce})$ within the plasma. 
In our studies we achieve this by choosing the following form of the magnetic field profile with the choice of $\delta = 0.02$. 
\begin{equation}
    \vec{B}_{ext}= [2.2 -0.02(x-110)]\hat{i}+ [0.02(y-110)]\hat{j} B_N
\end{equation}
Under the application of this profile of the  magnetic field, ECR condition is met  at the edge of the plasma lens. This has been shown in Figure \ref{fig:schematic_2}.  Here the red dot shows the ECR point where cyclotron frequency matches with EM wave frequency. 

The high gradient magnetic field  focuses the  wave pulse very rapidly and thus a RCP wave pulse with minimum spot size is achieved  at the plasma edge. The ECR condition then  results in the dumping of the EM wave energy to the electrons very rapidly.  This is consistent with one of our earlier studies   \citep{PhysRevE.110.065213} where we have shown the conversion of EM field energy to electron kinetic energy at the ECR point. This has also been  illustrated from figure \ref{fig:time_evolution+energy_absorption}. In subplot $(a)$ we depict the  EMF energy as a function of time and the transverse dimension $y$. We have averaged it over the $x$ dimension here. 
It should be noted that it maximizes when the pulse gets focused inside the plasma. However, at later stage the EMF energy gets weakened. This happens as the EMF energy gets transferred to the electron kinetic energy at the resonance point. This interdependence gets clearly depicted from  subplots $(b)$ and $(c)$. Here we have shown the EMF energy and the electron kinetic energy by color plots as a function of $x$ and $t$. It should be noted that the EMF energy gets weaker at the specific location $x = x_{ECR}$ which  is the resonance point. Exactly at this location the electrons acquire a huge kinetic energy. 

\begin{figure}
  \centering
  \includegraphics[width=6cm]{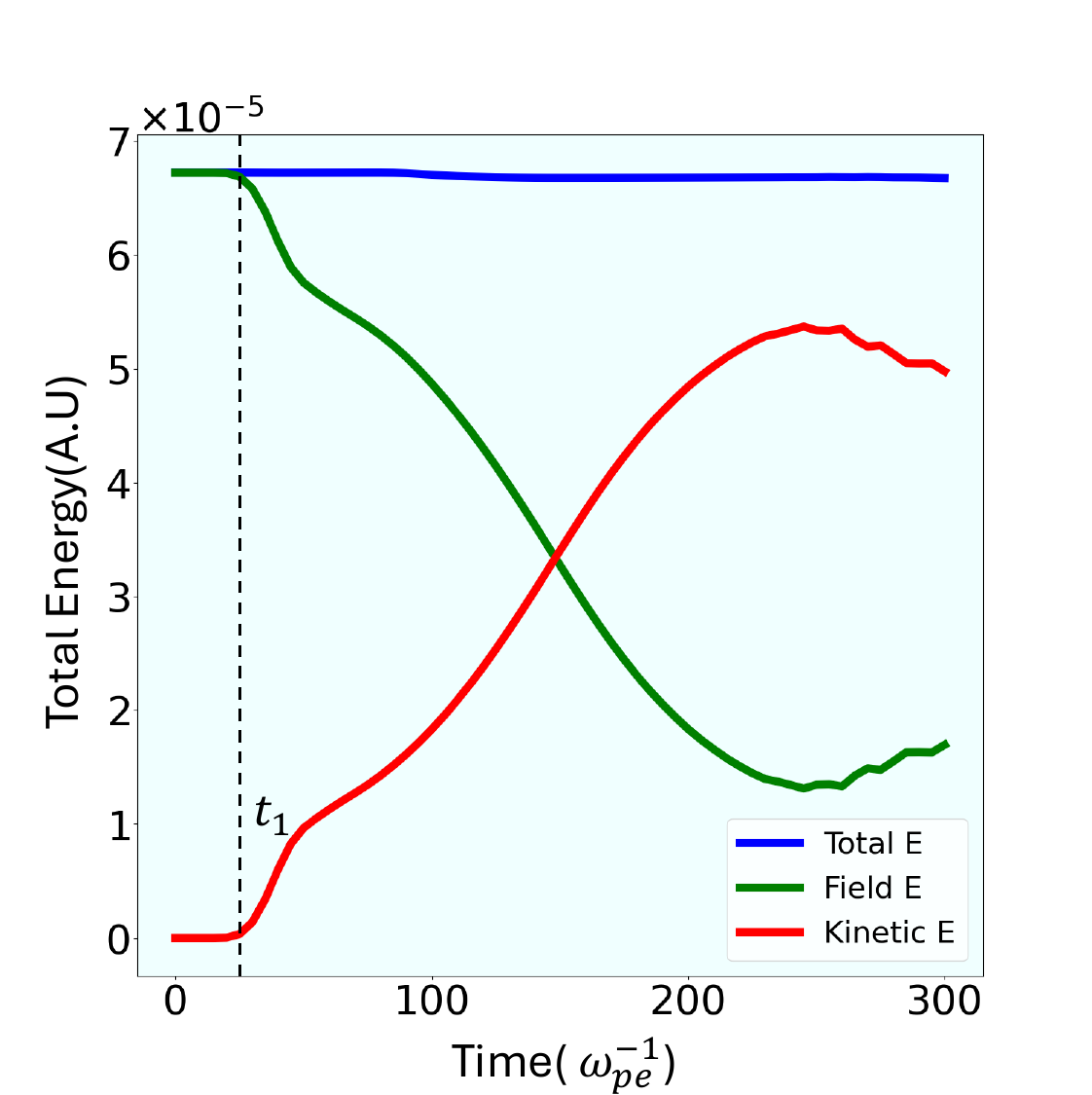}% Images in 100% size
  \caption{Time evolution of Total energy, EMF energy and kinetic energy of electrons for the case of high magnetic field gradient.}
\label{fig:Energy_lens_1}
\end{figure}

\begin{figure*}
  \centering
  \includegraphics[width=14cm]{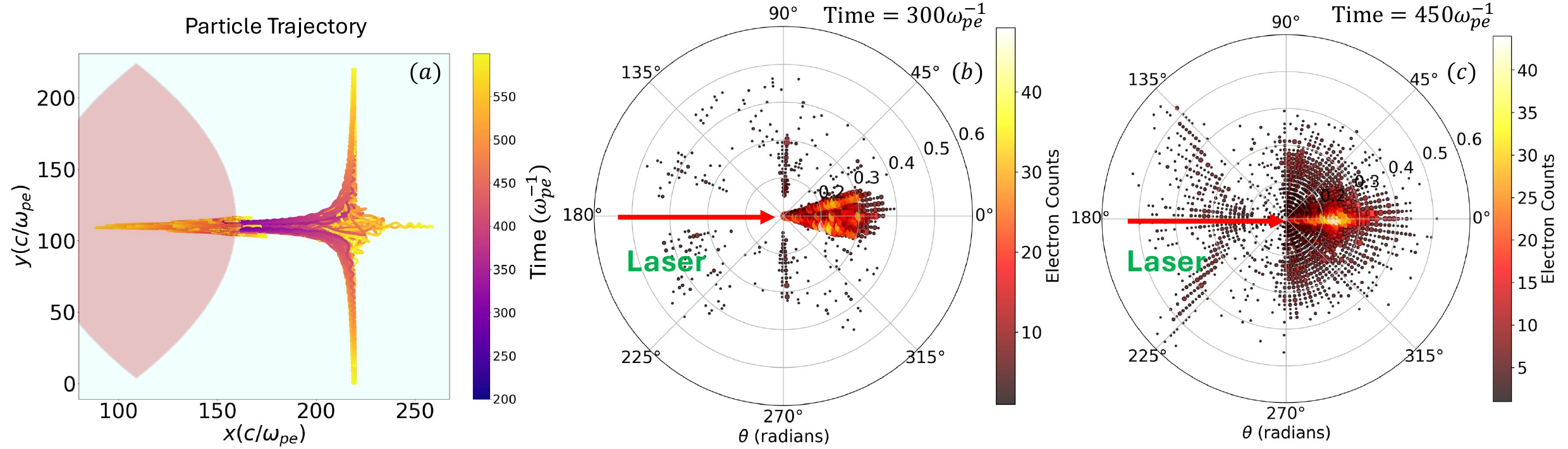}% Images in 100% size
  \caption{Figures illustrates a) the trajectory of randomly chosen 2000 electrons for  color axis with respect to time (until $600\omega_{pe}^{-1}$) and angular distribution of electrons at time b) $300\omega_{pe}^{-1}$ and c) $450\omega_{pe}^{-1}$.}
\label{fig:particle_trajectory_new}
\end{figure*}

Figure \ref{fig:Energy_lens_1} also shows the energy evolution with time. The graph shows that nearly $80\%$ of the pulse energy has been transferred to electrons. Particle trajectory of these randomly chosen 2000 electrons from focal spot location have been shown in figure \ref{fig:particle_trajectory_new}(a). Figure (a) shows the trajectory with time on the color axis. The transverse dimension of the emerging beam is about $5l_N$. These ejected particles strongly follow the external magnetic field direction and divides into two electron beams and propagates away. 

Figure \ref{fig:particle_trajectory_new}(b,c) shows, angular  distribution of these particle at time  $300 \omega_{pe}^{-1}$ and $450 \omega_{pe}^{-1}$ with color axis showing electron counts and radius of the circle represents  energy in MeV. At $300 \omega_{pe}^{-1}$ , these ejected electrons haven almost directional distribution and directed along the laser axis. Later in time they diverges and spreads along the perpendicular direction of laser axis due to the external magnetic field present in vacuum.  Maximum electron energy about $E\sim 0.78 m_ec^2 =0.4 MeV$ against the incident EM wave of normalized vector potential  $a_0=0.08$. \\

This is an efficient mechanism to produce highly focused electron beam through self-focusing of EM wave. Thus, appropriate tailored parameter can be used in experiments for collimated electron beam generation.  

%%%%%%%%%%%%%%%%%%%%%%%%%%%%%%%%%%%%%%%%%%%%%%%%%%%%%%%%%%%%%%%%%%%%%%%%%%%%%%%%%%%%%%%%%%%%%%%%%%%%%%%%%%%%%%%%%%%%%%%%%%
\section{Summary}\label{sec:summary}
    The above discussion shows that when a linearly polarized electromagnetic wave pulse propagates through a underdense homogeneous plasma lens embedded in a slowly diverging type of external magnetic field , it divides into RCP and LCP wave pulse and spot size of RCP wave pulse reduces to a minimum size through self-focusing. The electrons in the plasma under the influence of electromagnetic wave, drift along external magnetic field direction and causes a density compression and depression which modifies the refractive index of plasma and the EM wave focus along the propagation axis. Our analysis has shown that focusing is also dependent on shape of plasma. With just a $79\mu m$ thin shaped plasma lens $(c=0.005)$, magnetic field induced focusing reduces spot size of an incident RCP wave by almost $\sim 1/3.5$ and increases intensity by almost three times.  One of the main advantages of using an underdense magnetic lens is that energy loss in plasma could be minimal and a focused wave pulse can be obtained. Further, it is also established that by increasing the gradient of the magnetic field, a wave pulse can be focused inside the plasma at ECR resonance and can generate a strong collimated electron beam. The theory can find applications in the focusing of strong intense lasers and industrial plasma as well as particle accelerator schemes. Magnetic field-induced focusing can play a major role in inertial confinement fusion as well.   
%%%%%%%%%%%%%%%%%%%%%%%%%%%%%%%%%%%%%%%%%%%%%%%%%%%%%%%%%%%%%%%%%%%%%%%%%%%%%%%%%%%%%%%%%%%%%%%%%%%%%%%%%%%%%%%%%%%%%%%%%%

 \section*{Acknowledgements}
 \indent The authors would like to acknowledge the OSIRIS Consortium, consisting of UCLA and IST (Lisbon, Portugal), for providing access to the OSIRIS 4.0 framework, which is the work supported by the NSF ACI-1339893. A.D. acknowledges support from the Anusandhan National Research Foundation (ANRF) of the Government of India through core grant CRG/2022/002782 as well as her J C Bose Fellowship grant JCB/2017/000055. The authors would like to thank IIT Delhi HPC facility for computational resources. T.D. also wishes to thank the Council for Scientific and Industrial Research (Grant No. 09/086/(1489)/2021-EMR-I) for funding the research.

\section*{Conflict of Interest}
 \noindent Authors report no conflict of interest

\bigskip
\bigskip

\bibliographystyle{jpp}

\bibliography{jpp-instructions.bib}

\end{document}